\documentclass [12pt]{iopart}
\usepackage{epsfig}
\begin{document}

\title[Hyperspherical Coulomb spheroidal representation in
the three-body problem] {Hyperspherical Coulomb spheroidal
representation in the Coulomb three-body problem}

\author{D I Abramov}

\address{V A Fock Institute of Physics,
St. Petersburg State University, 198504, Russia}

\begin{abstract}
The new representation of the Coulomb three-body wave function via
the well-known solutions of the separable Coulomb two-centre
problem $\phi_j(\xi,\eta)=X_j(\xi)Y_j(\eta)$  is obtained, where
$X_j(\xi)$ and $Y_j(\eta)$ are the Coulomb spheroidal functions.
Its distinguishing characteristic is the coordination with the
boundary conditions of the scattering problem below the
three-particle breakup. That is, the wave function of the
scattering particles in any open channel is the asymptotics of the
single, corresponding to that channel, term of the expansion
suggested. The effect is achieved due to the new relation between
three internal coordinates of the three-body system and the
parameters of $\phi_j(\xi,\eta)$. It ensures the orthogonality of
$\phi_j(\xi,\eta)$ on the sphere of a constant hyperradius,
$\rho=const$, in place of the surface
$R=|\bi{x}_2-\bi{x}_1|=const$ appearing in the traditional
Born-Oppenheimer approach. The independent variables $\xi$ and
$\eta$ are the orthogonal coordinates on that sphere with three
poles in the coalescence points. They are connected with the
elliptic coordinates on the plane by means of the stereographic
projection. For the total angular momentum $J\ge 0$ the products
of $\phi_j$ and the Wigner $D$-functions form the hyperspherical
Coulomb spheroidal (HSCS) basis on the five-dimensional
hypersphere, $\rho$ being a parameter. The system of the
differential equations and the boundary conditions for the radial
functions $f^J_i(\rho)$, the coefficients of the HSCS
decomposition of the three-body wave function, are presented.
\end{abstract}

\maketitle

\section{Introduction}

The complete set of the solutions of the Coulomb two-centre (CTC)
problem is widely used as a basis for the representation of the
Coulomb three-body (CTB) wave function for a long time (\cite{VP}
and references therein). One of the most attractive features of
that basis is its simplicity. First, due to the high symmetry of
the Coulomb field, the CTC problem admits the separation of
variables, and the basis elements are the completely factorized
functions. Second, the co-factors composing the basis elements,
so-called Coulomb spheroidal functions (CSF), are the well
investigated functions related to the Heun class. The principal
results on CSF and the references are presented in
\cite{PWR}-\cite{SLAV}.

The traditional way to use the CTC basis for the analysis of the
CTB system with positive and negative charges ($Z_1Z_2>0$,
$Z_1Z_3<0$) is the adiabatic representation \cite{VP} which is
known also as the Born-Oppenheimer (BO) approach, the perturbed
stationary states method,  the molecular orbitals method. In that
approach the distance $R$ between the particles 1 and 2 is
considered as an adiabatic parameter, and the basis functions are
orthogonal at fixed $R$. The numerous helpful properties of CSF
allow to construct the effective numerical algorithms for the
calculation of bound states and cross sections of elastic and
inelastic collisions for various CTB systems. The method is
successfully used not only for the case of heavy particles 1,2 and
a light particle 3 \cite{VP}, \cite{MPF}-\cite{Kor2}, but as well
for light particles 1,2 (electrons) and a heavy particle 3
(nucleus) \cite{FB1}-\cite{ROST2}.

Nevertheless, the BO representation in the traditional form (i.e.
with $R$ as a parameter) has a well-known imperfection which
decreases the efficiency of the calculations of the scattering
processes. The point is that the BO expansion is not coordinated
with the physical boundary conditions of the scattering problem.
It means that at $R\to\infty$ the wave function of the scattering
particles in any open channel is formed as the sum of the infinite
number of components, the selected terms of the BO series
calculated at large $R$. That sum includes, in particular, the
basis functions tending at large $R$ to the atomic wave functions
of the closed channels. Thus, the terms of the BO expansion,
generally speaking, are not divided into the groups corresponding
to open and closed channels. The peculiarity pointed out becomes
apparent in the system of radial equations: the matrix elements
coupling different basis states may remain nonzero at large $R$.
That leads to the complication of the boundary conditions for the
radial functions at $R\to\infty$ \cite{Kor1} and to the slow
convergence of the expansion. The incoordination of the adiabatic
representation with the physical boundary conditions of the
three-body (not necessary Coulomb) scattering problem is peculiar
to both the quantum approach and the semiclassical one. To
overcome the difficulties in the frames of the semiclassical
treatment the inclusion of "electron translation factors" is used
\cite{ETF1}, \cite{ETF2}. The various methods of the improvement
of the  adiabatic representation for the scattering processes are
developed also in the quantum approach (\cite{Bel1}, \cite{Bel2}
and references therein).

In contrast to the traditional BO representation the adiabatic
hyperspherical (AHS) one \cite{Mac}-\cite{AGP1} is coordinated
with the boundary conditions of the three-body scattering problem.
In that approach  the hyperradius $\rho$ is used as an adiabatic
parameter. The wave function of any open channel is the
asymptotics of the single, corresponding to that channel, term of
the expansion. The AHS approach is successfully applied to the
calculations of the bound states, the cross sections of the
elastic and inelastic collisions and the resonances in various CTB
systems \cite{Mac}, \cite{Lin}-\cite{AGR2}.

However, the AHS basis elements are the essentially more
complicated functions than the BO ones. In the general case they
are the non-factorized functions of five independent variables. In
addition, for the Coulomb three-body systems the AHS energy terms
as functions of $\rho$ have the numerous avoiding crossings which
make difficulties for the numerical calculations \cite{Lin},
\cite{AGP1}. These avoiding crossings are closely connected with
the exact crossings of the CTC adiabatic terms \cite{AGP1}. Thus,
in the AHS approach the high symmetry of the Coulomb field,
instead of the simplification, paradoxically leads to the
additional difficulties. The avoiding crossings are removed in the
original version of the hyperspherical approach developed in
papers \cite{T1}-\cite{T5}. In these papers the basis completely
factorizing in the special four-pole elliptic coordinates on the
hypersphere is suggested. The co-factors composing the basis
elements in that approach are essentially more complicated
functions as compared with CSF, they relate to the Heun class only
at zero energy \cite{T4}.

The goal of the present paper is to obtain the representation of
the CTB  wave function which, on the one hand, contains only the
well-known CSF and, on the other hand, is coordinated with the
boundary conditions of the scattering problem  at energies below
the threshold of the three-particle breakup. In other words, to
obtain the representation combining the advantages of the AHS
approach (the coordination with the boundary conditions) and the
advantages connected with the specific character of the Coulomb
field (the complete factorization of the basis and the simplicity
of its components, CSF). That task is interesting, firstly, for
the development of the efficient computing methods and, secondly,
for the completeness of the theoretical description of CSF:  it is
desirable to present various ways to use these well-known
functions in the CTB problem.

The above goal is achieved in the paper with the help of the new
relation between the parameters of CSF and the internal
coordinates of the CTB system. The suggested CTC basis functions,
just as the AHS basis functions, form the orthogonal set on the
sphere $\rho=const$ in the three-dimensional space of the internal
variables, in place of the surface $R=const$ appearing in the
traditional BO approach.

The difference of our expansion from the traditional BO one can be
outlined briefly for the simplest case of zero total angular
momentum, $J=0$, in the following way. Let $\bi{x}_1$, $\bi{x}_2$,
and $\bi{x}_3$ be the position vectors of the particles in the
centre-of-mass frame
\begin{equation}
m_1\bi{x}_1+m_2\bi{x}_2+m_3\bi{x}_3=0. \label{cm}
\end{equation}
In the case $J=0$ the three-body wave function $\Psi$ depends on
three independent internal variables, for example $r_1$, $r_2$,
$R$:
\begin{equation}
{r}_1=|\bi{x}_3-\bi{x}_1|,\quad{r}_2=|\bi{x}_3-\bi{x}_2|,\quad
{R}=|\bi{x}_2-\bi{x}_1|. \label{Intr}
\end{equation}

In the traditional BO approach three independent variables are, in
fact, $R$ and the prolate spheroidal coordinates
$\xi\in[1,\infty)$ and $\eta\in[-1,1]$ defined by the relations
\begin{equation}
\xi=(r_1+r_2)/R,\qquad \eta=(r_1-r_2)/R.
\label{Intxi}\end{equation} The variable $R$ is considered as a
parameter of the basis, and the decomposition of the three-body
wave function has the form
\begin{equation}
\Psi=\sum_j\tilde f_j(R)\tilde\phi_j(R|\xi,\eta),
\end{equation}
where the basis functions $\tilde \phi_j(R|\xi,\eta)$ are the
factorized solutions of the CTC problem with the intercentre
distance $R$. These solutions as the functions of $\xi$ and $\eta$
form the complete orthogonal system at fixed $R$.

In our approach the independent variables are $\xi$, $\eta$ and
the hyperradius
\begin{eqnarray}\fl
\rho= \sqrt{2(m_1{x}^2_1+m_2{x}^2_2+m_3{x}^2_3)}\nonumber\\
\lo=\sqrt{2(m_1m_3r_1^2+m_2m_3r_2^2+m_1m_2R^2)/(m_1+m_2+m_3)},
\quad \rho\in[0,\infty). \label{IntRho}
\end{eqnarray}
It is considered as the parameter of the basis.  The decomposition
of the three-body wave function has the form
\begin{equation}
\Psi=C(\rho,\xi,\eta)\sum_j f_j(\rho) \phi_j(\rho|\xi,\eta),
\end{equation}
where $C(\rho,\xi,\eta)$ is the special weight factor. Basis
functions $\phi_j(\rho|\xi,\eta)$ are the factorized solutions of
the CTC problem with the modified charges and the intercentre
distance proportional to the hyperradius $\rho$. They form the
complete orthogonal system at fixed $\rho$ and coincide (up to a
constant factor) with $\tilde \phi_j(R|\xi,\eta)$ at large $R$.

In the case $J>0$ the basis suggested consists of the productions
of CSF and the symmetrized Wigner functions. We call it "the
hyperspherical Coulomb spheroidal (HSCS) basis". The formulae are
more complicated but the principal idea is the same.

The principal notations and the starting equations are presented
in section 2. Section 3 is devoted to the construction of the HSCS
basis. In section 4 the definition of $S$-matrix in the
appropriate representation is presented, and the asymptotical
expressions for the radial functions in terms of the matrix
elements of $S$-matrix are obtained. These formulae are deduced
from the properties of the HSCS basis and the general relations
defining $S$-matrix, without any analysis of the system of the
radial equations. That system and the statement of the scattering
problem at energies below the three-particle breakup are presented
in section 5.

\section{Starting equations}

We consider the system of three particles with charges $Z_1>0$,
$Z_2>0$, $Z_3=-1$ and masses $m_1$, $m_2$, $m_3$. The case of the
identical particles 1 and 2 is not considered. The atomic units
are used. The Jacobi coordinates $\bi{R},\bi{r}$ are expressed in
terms of the position vectors of the particles  $\bi{x}_1$,
$\bi{x}_2$, $\bi{x}_3$ \eref{cm} by the formulae
\begin{equation}
\bi{R} =\bi{x}_2-\bi{x}_1,\qquad\bi{r}
=\bi{x}_3-\frac{m_1\bi{x}_1+m_2\bi{x}_2}{m_1+m_2}, \label{e1}
\end{equation}
where $M$ and $\mu$ are the reduced masses:
\begin{equation}
M=\frac{m_1m_2}{m_1+m_2},\qquad \mu =\frac{m_3(m_1+m_2)}
{m_1+m_2+m_3}. \label{e4}\end{equation} The Jacobi coordinates
$\bi{r}_{\alpha},\bi{R}_{\alpha}$, which are suitable at large $R$
for the clusterization $\alpha$ ($\alpha=1$ corresponds to the
atom (1,3) plus the distant particle 2; $\alpha=2$ corresponds to
(2,3)+1), are defined by
\begin{equation}
\bi{r}_{\alpha}=\bi{x}_3-\bi{x}_{\alpha},\qquad
\bi{R}_{\alpha}=(-1)^{\alpha-1}\left[\bi{x}_{3-\alpha}
-\frac{m_{\alpha} \bi{x}_{\alpha}+
m_3\bi{x}_3}{m_{\alpha}+m_3}\right], \label{rrrr}
\end{equation}
\begin{equation}
\mu_{\alpha}=\frac{m_{\alpha}m_3}{m_{\alpha}+m_3},\qquad
M_{\alpha}=\frac{m_{3-\alpha}(m_{\alpha}+m_3)}{m_1+m_2+m_3},
\qquad \alpha=1,2. \label{mm}\end{equation} The factor
$(-1)^{\alpha-1}$ \eref{rrrr} ensures the identical directions of
$\bi{R}_{\alpha}$ and $\bi{R}$ at large $R$.

The Hamiltonian of the system in the coordinates $\bi{R},\bi{r}$
has the form
\begin{equation}
H=T+V,\label{H}\end{equation}\begin{equation} T
=-\frac{1}{2M}\Delta_R-\frac{1}{2\mu}\Delta_r, \label{T}
\end{equation}
\begin{equation}
 V=
\frac{Z_1Z_2}{R}-\frac{Z_1}{|\bi{r}+\bi{R}M/m_1|}
-\frac{Z_2}{|\bi{r}-\bi{R}M/m_2|}.\label{V}\end{equation}

In the hyperspherical approach six variables
\begin{equation}\fl \Phi\in[0,2\pi),\quad\Theta\in[0,\pi],\quad
\varphi\in[0,2\pi),\quad
\rho\in[0,\infty),\quad\chi\in[0,\pi],\quad\vartheta\in[0,\pi]
\label{e51}\end{equation} are generally used. Here
$\{\Phi,\Theta,0\}$ are the Euler angles of the rotating frame
with the third axis directed along $\bi{R}$. The variables
$\vartheta$ and $\varphi$ are the spherical angles of $\bi{r}$ in
the rotating frame. The hyperradius $\rho$ and the hyperangle
$\chi$ are  expressed in terms of ${ R}$ and ${ r}$ as
\begin{equation}
\rho^2=2(MR^2+\mu r^2),\qquad \rho\in[0,\infty);\label{e7}
\end{equation}
\begin{equation}
\tan\chi/2=(\mu/M)^{1/2}r/R,\qquad \chi\in[0,\pi].
\label{e8}\end{equation} The angles $\Phi$, $\Theta$, $\varphi$
are the external coordinates defining the orientation of the
triangle formed by three particles. The variables $\rho$, $\chi$,
$\vartheta$ are the internal coordinates of the system. In the
hyperspherical coordinates the kinetic energy \eref{T}  takes the
form \cite{KV,AGP1}
\begin{equation} T= -{{1}\over{\rho^5}}{{\partial }\over{\partial
\rho}} \rho^5 {{\partial}\over{\partial \rho}} -
{{4}\over{\rho^2\sin^2 \chi}} \biggl( {{\partial }\over{\partial
\chi}} \sin^2\chi {{\partial }\over{\partial \chi}} -{\bi{l}}^2
\biggr) + {{{\bi{ J}}^2 - 2 ({\bi{ l}}, {\bi{ J}})}\over{\rho^2
\cos^2 \chi/2}}. \label{Th}\end{equation} The operators $\bi{J}$
and $\bi{l}$ in \eref{Th} are the total angular momentum and the
angular momentum of the third particle respectively. The
expressions for $\bi{l}^2$, $\bi{J}^2$, and $\bi{J}\cdot\bi{l}$
are given by

\begin{eqnarray}
\bi{l}^2= -\frac{1}{\sin\vartheta}{{\partial }\over{\partial
\vartheta}} \sin \vartheta {{\partial }\over{\partial \vartheta}}
-\frac{1}{\sin^2\vartheta} {{\partial^2 }\over{\partial
\varphi^2}},\label{e14}\\
\bi{J}^2 = -{{1}\over{\sin \Theta}} {{\partial }\over{\partial
\Theta}} \sin \Theta {{\partial }\over{\partial \Theta}}
-{{1}\over{\sin^2 \Theta}} \biggl( {{\partial^2 }\over{\partial
\Phi^2}} + {{\partial^2 }\over{\partial \varphi^2}} - 2\cos \Theta
{{\partial^2 }\over{\partial \Phi\partial\varphi}} \biggr),
\label{e15}\\
\bi{J}\cdot\bi{l} = \left[ {{\sin \varphi}\over{\sin \Theta}}
\bigl( \cos \Theta {{\partial }\over{\partial \varphi}}-
{{\partial }\over{\partial \Phi}} \bigr) - \cos  \varphi
{{\partial }\over{\partial \Theta}} \right]{{\partial
}\over{\partial \vartheta}}+\nonumber\\
+ \left[ \sin \varphi {{\partial^2 }\over{\partial \Theta
\partial \varphi}}
- {{\cos \varphi}\over{\sin \Theta}} {{\partial^2 }\over{\partial
\Phi
\partial \varphi}}  +        \cos \varphi
\cot \Theta {{\partial^2 }\over{\partial \varphi^2}} \right]
\cot\vartheta - {{\partial^2 }\over{\partial \varphi  ^2}}.
\label{e16}\end{eqnarray}

In what follows we use instead of $\chi$ the variable
\begin{equation}t=\tan{\chi/2}, \qquad t\in[0,\infty).
\label{t}\end{equation} One can consider $t$ and $\vartheta$ as
the polar coordinates on the half-plane presenting the
stereographic projection of the hemisphere with the radius 1, the
latitude $\chi$ and the longitude $\vartheta$ on the equatorial
plane (figure 1). Three independent variables $t$, $\vartheta$ and
$\varphi$ are the spherical coordinates of the vector $\bi{t}$ in
the rotating frame. That vector differs from the relative vector
$\bi{r}/R$ \cite{SV} by the mass factor:
\begin{equation} \bi{t}= t\bi{r}/r=(\mu/M)^{1/2}\bi{r}/R.
\label{tvec}\end{equation}

For the formulation of our approach it is suitable to write the
three-body Hamiltonian $H$ \eref{H}, \eref{T},
\eref{V} 
in terms of the variables $\Phi, \Theta, \varphi, \rho, t,
\vartheta $. Taking into account the expression for  the six
dimensional elementary volume
\begin{equation}\rmd\bi{R}\rmd\bi{r}=(4M\mu)^{-3/2}gt^2\sin{\vartheta}
\sin{\Theta}\rmd \rho\rmd
t\rmd\vartheta\rmd\varphi\rmd\Theta\rmd\Phi,
\label{dV6}\end{equation}
\begin{equation}g=\rho^5(1+t^2)^{-3}, \label{g}
\end{equation}
one can present the kinetic energy $T$ \eref{T} as
\begin{equation}T=g^{-1/2}\tilde T g^{1/2},
\label{Tt}\end{equation}
\begin{equation}
\tilde T=-\frac{\partial^2}{\partial\rho^2}+\frac{1}{\rho^2}
\left[
-(1+t^2)\Delta_t(1+t^2)+(1+t^2)(\bi{J}^2-2\bi{J}\cdot\bi{l})
+\frac{3}{4}\right],\label{tT}\end{equation} where
\begin{equation}
 \Delta_t=\frac{1}{t^2}\frac{\partial}{\partial t}t^2
\frac{\partial}{\partial t}- {{{\bf
l}^2}\over{t^2}}.\label{Dt}\end{equation}

Using \eref{tvec} and the notations
\begin{equation}
\bi{t}_{\alpha}=(-1)^{\alpha}t_{\alpha}\bi{R}/R, \qquad
t_{\alpha}=(\mu M)^{1/2}m_{\alpha}^{-1},\qquad
{\alpha}=1,2,\label{tj}\end{equation} we obtain $V$ \eref{V} in
the form
\begin{equation}
V= \frac{(1+t^2)^{1/2}}{\rho}\left[(2M)^{1/2}Z_1Z_2-
\frac{(2\mu)^{1/2}Z_1}{|{\bi{ t}}-{\bi
{t}}_1|}-\frac{(2\mu)^{1/2}Z_2}{|{\bi {t}}-{\bi{t}}_2|}\right],
\label{Vt}\end{equation}
\begin{equation}|{\bi{t}}-{\bi{t}}_{\alpha}|
=\left[t^2+t_{\alpha}^2-
2(-1)^{\alpha}tt_{\alpha}\cos{\vartheta}\right]^{1/2},\qquad
{\alpha}=1,2.\label{t-tj}
\end{equation}

The three-body wave function
$\Psi=\Psi(\rho,t,\vartheta,\Phi,\Theta,\varphi)$ satisfies the
Schr\"{o}dinger equation
\begin{equation}H\Psi=E\Psi.\label{Schr}\end{equation} Using
the transformation \eref{Tt} one can present it in the form
\begin{equation}
\tilde H\tilde\Psi=E\tilde\Psi, \label{tSchr}\end{equation} where
the transformed wave function $\tilde \Psi$ and the transformed
three-body Hamiltonian $\tilde H$ are given by the relations
\begin{equation}
\tilde\Psi=g^{1/2}\Psi=\rho^{5/2} (1+t^2)^{-3/2} \Psi,
\label{tPsi}\end{equation}
\begin{equation}
\tilde H=g^{1/2}Hg^{-1/2}= -\frac{\partial^2}{\partial\rho^2}+
\tilde h^{\rm a\rmd}+\frac{(1+t^2)}{\rho^2} \left(\bi{J}^2-
2\bi{J}\cdot\bi{l} \right)+ \frac{3}{4\rho^2},
\label{tH}\end{equation}
\begin{equation}
\tilde h^{\rm a\rmd}=-\rho^{-2}(1+t^2)\Delta_t(1+t^2)+V.
\label{had}\end{equation}

The transformed AHS Hamiltonian $\tilde h^{\rm a\rmd}$ \eref{had}
acts on functions of $t,\vartheta,\varphi$ and depends on $\rho$
parametrically. Its eigenfunctions are used as a basis for the
representation of the three-body wave function in the reduced
version of the AHS approach \cite{AGP1}, \cite{AGP2}. (In the
general version \cite{KV}, \cite{Lin} the AHS Hamiltonian $h^{\rm
a\rmd}$ includes two last terms of (26) and acts on functions of
five variables.)

\section{The hyperspherical Coulomb spheroidal (HSCS) basis}

\subsection{The appropriate Coulomb two-centre (CTC) Hamiltonian in {\bf
t}-space}

It is seen that $\tilde h^{\rm a\rmd}$ \eref{had} can be
considered as the Hamiltonian of the particle with the variable
mass $\tilde m(t)$ moving in the field $V$ in ${\bi t}$-space:
\begin{equation}
\tilde h^{\rm a\rmd}=-(2\tilde m)^{-1/2}\Delta_t(2\tilde
 m)^{-1/2}+V,\qquad 2\tilde m(t)=\rho^2(1+t^2)^{-2}.
\label{had1}\end{equation} Here the form of the kinetic energy
ensures the self-conjugacy. The pure discrete character of the
spectrum of $\tilde h^{\rm a\rmd}$ is explained by the fast
decrease of the mass $\tilde m(t)$ at large $t$, $\tilde m(t)\sim
t^{-4}$. Every eigenfunction of $\tilde h^{\rm a\rmd}$ coincides
at large $\rho$ with the wave function of certain bound state of
the atom (1,3) or (2,3) \cite{AGP1}.

For the generating of the basis functions in our approach we also
use the Hamiltonian of the particle moving in ${\bi t}$-space, but
the mass of the particle is fixed. Namely, we use the traditional
CTC Hamiltonian
\begin{equation}
h=-\Delta_t-\frac{\tilde Z_1}{|\bi{t}-\bi{t}_1|}-\frac{\tilde Z_2}
 {|\bi{t}-\bi{t}_2|}.\label{h} \end{equation}
It describes the motion of the particle with the mass 1/2 and the
charge -1 in the field of two effective charges $\tilde Z_1$ and
$\tilde Z_2$ located in the points $\bi{t}_1$ and $\bi{t}_2$ of
$\bi{t}$-space at the distance
\begin{equation} |\bi{t}_2-\bi{t}_1|=
t_1+t_2=(\mu/M)^{1/2}\label{t1+t2}\end{equation} from each other.
In our approach the effective charges $\tilde Z_1=\tilde
Z_1(\rho)$ and $\tilde Z_2=\tilde Z_2(\rho)$ are chosen so that
every eigenfunction of the discrete spectrum of $h$ \eref{h} at
large $\rho$ coincides, as well as in the case of $\tilde h^{\rm
a\rmd}$, with the wave function of some bound state of the atom
(1,3) or (2,3). The simplest choice is the linear function $\tilde
Z_j(\rho)$ for which we easy obtain
\begin{equation}\tilde Z_{\alpha}=\rho Z_{\alpha}(2\mu)^{1/2}
(1+t_{\alpha}^2)^{-3/2}= \rho
Z_{\alpha}{{2}^{1/2}\mu_{\alpha}^{3/2}} {\mu}^{-1},\qquad
{\alpha}=1,2. \label{tildeZ}
\end{equation}
Indeed, under that condition each eigenfunction of the discrete
spectrum of $h$ \eref{h} at $\rho\to\infty$ is localized (with the
exception of the case $\tilde Z_1=\tilde Z_2$, see below) in the
neighborhood of one of the centres and coincides with the
eigenfunction of the one-centre Hamiltonian \cite{KPS}. For
example, in the asymptotic region
\begin{equation} R\to\infty,\qquad |\bi{x}_3-\bi{x}_1| = r_1=O(1),
\label{Rto}
\end{equation} or, in the hyperspherical variables,
\begin{equation}\rho\rightarrow\infty,\quad|\bi{t}-\bi{t}_1|=
(\mu/M)^{1/2}r_1/R=r_1(2/\mu_1)^{1/2}\mu/\rho+O(\rho^{-2}),
\label{rhoto}
\end{equation}
the corresponding one-centre Hamiltonian is given by
\begin{equation}-\Delta_t-\frac{\tilde Z_1}{|\bi{t}-\bi{t}_1|}\approx
\frac{\rho^2\mu_{1}^2}{\mu^2}
\left(-\frac{1}{2\mu_{1}}\Delta_{r_{1}}-\frac{ Z_1}{r_{1}}\right).
\label{Delta}\end{equation} The expression in the parenthesis
coincides  with the Hamiltonian of the atom (1,3). Hence, the
eigenfunction of $h$ \eref{h} localized at $\rho\to\infty$ in the
domain \eref{Rto}, \eref{rhoto} coincides up to a constant factor
with the wave function of some bound state of the atom (1,3). The
indices of the state are unimportant here.

The formula for $\tilde Z_2$ is proved in exactly the same way.

As it is seen from equation \eref{tildeZ} in the case
\begin{equation}
Z_1/Z_2=(\mu_2/\mu_1)^{3/2} \label{ZZ}
\end{equation}
the effective charges are equal, $\tilde Z_1=\tilde Z_2$. In that
case the Hamiltonian \eref{h} has an additional symmetry, and its
eigenfunctions are divided into symmetric and antisymmetric ones.
Thus, every eigenfunction is localized at $\rho\to\infty$  near
both centres. For the construction of the basis in that case it is
suitable to use the corresponding linear combinations of symmetric
and antisymmetric eigenfunctions which are localized at large
$\rho$ in the vicinity of one of the centres. That transformation
does not present any difficulties. Taking into account the
abstract character of the condition \eref{ZZ} for the
non-identical particles 1 and 2 we do not consider that case and
suppose hereinafter
\begin{equation}
\tilde Z_1\neq\tilde Z_2. \label{tildZZ}
\end{equation}

\subsection{Coulomb spheroidal functions on the sphere $\rho=const$}

The equation defining the spectrum of the Hamiltonian
$h(\rho|\bi{t})$ \eref{h},\begin{equation}
\left(h-\varepsilon\right)\psi=0,\label{e28}\end{equation} is
investigated in details \cite{{PWR},{KPS}}. The problem admits the
separation of variables in the prolate spheroidal coordinates
$\xi\in[1,\infty)$ and $\eta\in[-1,1]$,
\begin{eqnarray}
\xi=|\bi{t}_1-\bi{t}_2|^{-1}\left(|\bi{t}-\bi{t}_1|+|\bi{t}-
\bi{t}_2|\right)= R^{-1}\left(r_1+r_2\right),
\label{xi} \\
\eta=|\bi{t}_1-\bi{t}_2|^{-1}\left(|\bi{t}-\bi{t}_1|-|\bi{t}-
\bi{t}_2|\right)=R^{-1}\left(r_1-r_2\right). \label{eta}
\end{eqnarray}

The eigenfunction  $\psi^{\pm}_{jm}(\rho|\bi{t})$, which
corresponds to the eigenvalue $\varepsilon=\varepsilon_{jm}<0$ in
the discrete spectrum and to $\varepsilon>0$ in the continuous
spectrum, is presented as the product
\begin{equation}
\psi^{\pm}_{jm}(\rho|\bi{t})= \phi_{jm}(\rho|\xi,\eta)e^{\pm
im\varphi}/\sqrt{2\pi}, \label{psi}
\end{equation}\begin{equation} \phi_{jm}(\rho|\xi,\eta)=
X_{jm}(\xi)Y_{jm}(\eta).\label{phi}
\end{equation} Here $m=0,1,2,...$ is the absolute value of the projection
of the total momentum on the rotating axis. The multi-index $j$ is
defined for the discrete spectrum $(\varepsilon<0)$ and for the
continuous one $(\varepsilon>0)$ by the equation
\begin{equation}
j=\cases{(n_{\xi},n_{\eta}), \quad \varepsilon<0,\cr
(\varepsilon,n_{\eta}), \quad
\varepsilon>0,\cr}\label{32}\end{equation} where $n_{\xi}$ and
$n_{\eta}$ are the number of nodes of radial $X(\xi)$ and angle
$Y(\eta)$ functions respectively. These functions, the so-called
Coulomb spheroidal functions (CSF), satisfy the system of
equations
\begin{eqnarray}
\left[\frac{\rmd}{\rmd\xi}(\xi^2-1)\frac{\rmd}{\rmd\xi}-\lambda-
p^2(\xi^2-1)+ a\xi\right]X(\xi)=0, \label{X}\\
\left[\frac{\rmd}{\rmd\eta}(1-\eta^2)\frac{\rmd}{\rmd\eta}+\lambda
-p^2(1-\eta^2)+b\eta\right] Y(\eta)=0 \label{Y}
\end{eqnarray}
and corresponding boundary conditions \cite{KPS}. Here $\lambda$
is the separation constant. The parameters $a$, $b$ and $p$ are
connected with the parameters of the Hamiltonian  $h$ \eref{h} by
the relations
\begin{eqnarray} a=\frac{1}{2}(\tilde Z_1+\tilde
Z_2)(t_1+t_2)=\frac{\rho}{\sqrt{2\mu M}}(Z_1\mu_1^{3/2}+
Z_2\mu_2^{3/2}),\label{a}\\
b=\frac{1}{2}(\tilde Z_2-\tilde Z_1)(t_1+t_2)=
\frac{\rho}{\sqrt{2\mu M}}(Z_2\mu_2^{3/2}-
Z_1\mu_1^{3/2}) , \label{b}\\
p=\frac{1}{2}(t_1+t_2)\sqrt{-\varepsilon}=\frac{1}{2}\sqrt
{-\frac{\mu\varepsilon}{M}}.\label{p}\end{eqnarray}

CSF defined by \eref{X}-\eref{p} differ from traditional CSF
\cite{KPS}, \cite{VP} by the new connection of the parameters $a$
and $b$ with the coordinates and the masses of three particles
\eref{a}, \eref{b}. The most important point is that our $a$ and
$b$ are proportional to $\rho$ instead of $R$ in the traditional
approach.

The functions $\phi_{jm}$ \eref{phi} at fixed $m$ and all possible
$j$ form the complete set in the space of functions of two
internal coordinates (\{$\chi,\vartheta$\},  \{$t,\vartheta$\} or
\{$\xi,\eta$\}) at fixed third internal coordinate, $\rho=const$.
The normalization accepted in the paper is defined by the equation
\begin{equation}
<\phi_{jm}|\phi_{j'm}>=\cases{\delta_{n_{\eta}n'_{\eta}}
\delta_{n_{\xi}n'_{\xi}}, \quad \varepsilon<0,\cr
\delta_{n_{\eta}n'_{\eta}}\delta(\varepsilon-\varepsilon'), \quad
\varepsilon>0,\cr}\label{norma}\end{equation} where $<f|g>$
denotes the two-dimensional integral
\begin{eqnarray}
\fl<f|g> \equiv \frac{1}{2}\int\limits_0^\pi
\frac{\sin^2{\chi/2}}{\cos^4{\chi/2}}\rmd \chi \int\limits_0^\pi
\rmd\vartheta f^*g\nonumber\\
=\int\limits_0^\infty t^2\rmd t \int\limits_0^\pi\sin\vartheta
\rmd\vartheta ~ f^*g =
\frac{(t_1+t_2)^3}{8}\int\limits_1^{\infty}\rmd\xi
\int\limits_{-1}^{1}\rmd\eta(\xi^2-\eta^2)~ f^*g.
\label{38}\end{eqnarray}

The completeness condition takes the form
\begin{eqnarray}
\sum_j \phi_{jm}(\rho|\xi,\eta)\phi_{jm}(\rho|\xi',\eta')=
\nonumber\\
\fl=\sum_{n_{\eta}=0}^{\infty}\left(\sum_{n_{\xi}=0}^{\infty}
\phi_{n_{\xi}n_{\eta}m}(\rho|\xi,\eta)\phi_{n_{\xi}n_{\eta}m}
(\rho|\xi',\eta') +\int\limits_0^{\infty} \phi_{\varepsilon
n_{\eta}m}(\rho|\xi,\eta)\phi_{\varepsilon
n_{\eta}m}(\rho|\xi',\eta')\rmd\varepsilon \right)= \nonumber\\
=8(t_1+t_2)^{-3}(\xi^2-\eta^2)^{-1}\delta(\xi-\xi')
\delta(\eta-\eta').\label{compl} \end{eqnarray}

It is convenient to consider the variables $\rho$, $\chi$ and
$\vartheta$ as  the spherical coordinates (the radius, the
latitude and the longitude respectively) in the three-dimensional
space with Cartesian coordinates $x=\rho\sin\chi\cos\vartheta$,
$y=\rho\sin\chi\sin\vartheta$, $z=\rho\cos\chi$. The net
$\xi=const$, $\eta=const$ on the sphere $\rho=const$ in that space
is presented on figure 1 (that figure is taken from the
monograph\cite{BOLS}, only the axes with the notations
corresponding to our problem are added). It can be obtained by the
stereographic projection of the sphere on the plane $z=0$ as a
preimage of the traditional orthogonal elliptic net on the plane.
Three poles correspond to the coalescence points of the three-body
system. As the stereographic projection is the conformal mapping
the net on the sphere is the orthogonal one.

\begin{figure}[h!]

\centerline{\scalebox{0.12}{\includegraphics{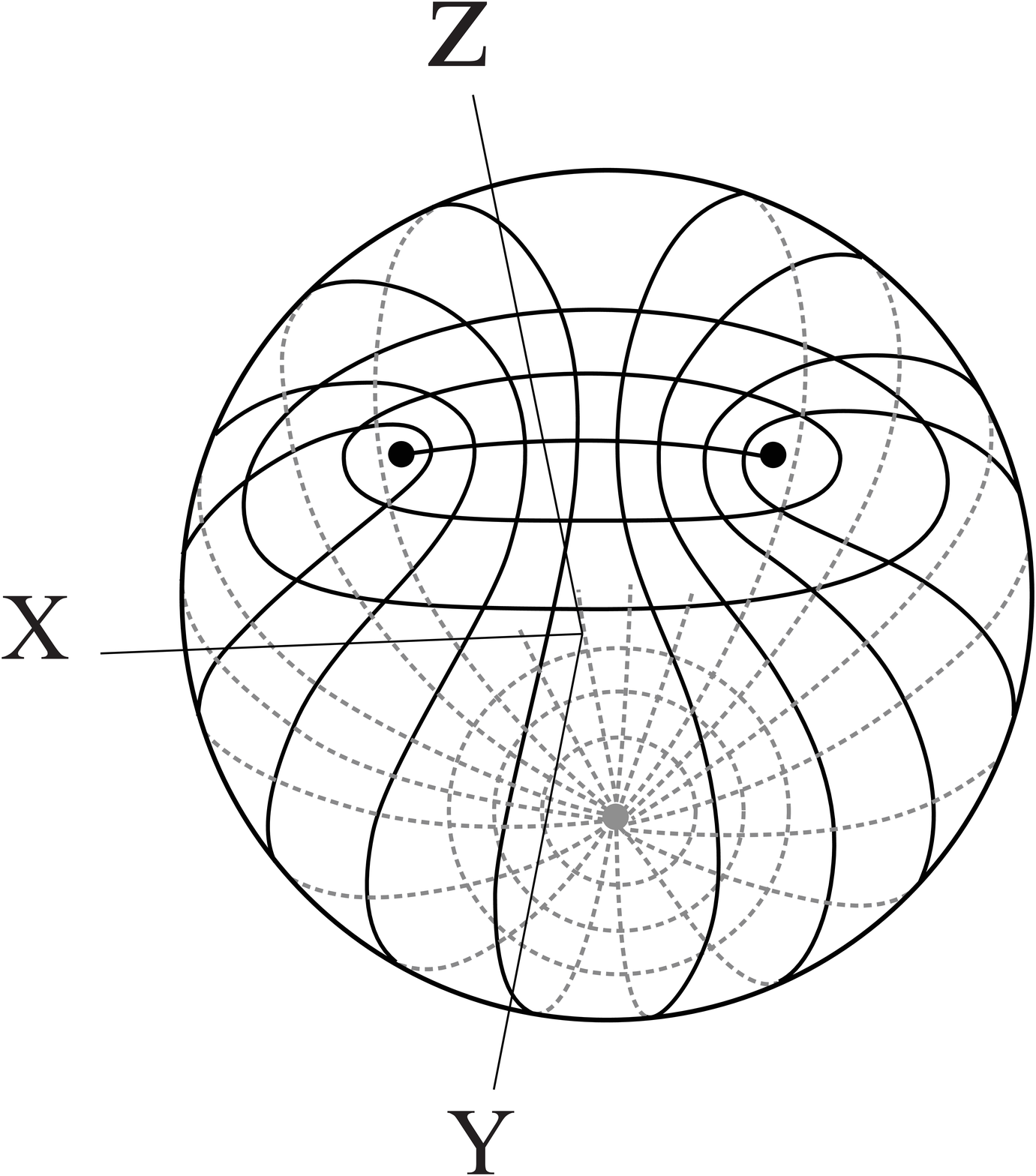}}} \caption {
The three-pole orthogonal net $\xi=const$, $\eta=const$ on the
sphere $\rho=\rho_0=const$ in the three-dimensional space
$x=\rho\sin\chi\cos\vartheta$, $y=\rho\sin\chi\sin\vartheta$,
$z=\rho\cos\chi$; $y\ge 0$ in our problem;  $\xi$ and $\eta$ are
given by \eref{xi}, \eref{eta}. The poles correspond to the
coalescence points of the three-body system. The stereographic
projection with the centre $\chi=\pi$, $z=-\rho_0$ onto the plane
$z=0$ gives the usual elliptic net on that plane. For $\rho_0=1$
the polar coordinates on the plane $z=0$ coincide with the
independent variables $t$ and $\vartheta$ of the CTC Hamiltonian
\eref{h}. }\end{figure}

\subsection{HCS basis on the five-dimensional hypersphere}

The three-body Hamiltonian \eref{H} has three commuting  integrals
of motion \cite{VP}: the square of the total angular momentum
$\bi{J}^2$, its projection on the third axis of the laboratory
frame $J_3$ and the inversion of all Jacobi coordinates $P$. It is
suitable therefore to use for the construction of the basis the
symmetrized Wigner D-functions $D_{Km}^{J\lambda}(\Phi, \Theta,
\varphi )$ which are the eigenfunctions of these operators. The
functions $D_{Km}^{J\lambda}(\Phi, \Theta, \varphi )$ are, in
addition, the eigenfunctions of the operator $J'^2_3$ where $J'_3$
is the projection of the total angular momentum on the third axis
of the rotating frame:
\begin{equation}
\bi{J}^2 D_{K m}^{J\lambda} = J(J+1) D_{K m}^{J\lambda}, \quad J_3
D_{K m}^{J\lambda} = K  D_{Km}^{J\lambda},
\end{equation}
\begin{equation}
 J'^2_3 D_{K m}^{J\lambda} = m^2 D_{Km}^{J\lambda}, \quad
PD_{Km}^{J\lambda} = \lambda  D_{K m}^{J\lambda},\quad \lambda=
\pm 1.\end{equation} The normalized symmetrized D-functions
$D_{Km}^{J\lambda}(\Phi, \Theta, \varphi )$ is given by the
expression \cite{AGP1}
\begin{equation}
D_{K m}^{J\lambda} =A^J_{Km}\left[ D_{-K m}^{~J} +\lambda
(-1)^{J+m} D_{-K -m}^{~J}\right], \label{D}\end{equation} where
\begin{equation}A^J_{Km}=(-1)^{K}
\left[(2J+1)/(1+
\delta_{0m})\right]^{1/2}/(4\pi)\label{A}.\end{equation} The
standard Wigner functions $D_{K m}^{J}(\Phi, \Theta, \varphi )$
are defined via  functions $d_{K m}^{J}(\Theta)$ \cite{Varsh}:
\begin{equation}
D_{K m}^{J}(\Phi, \Theta, \varphi )=\exp{(-iK\Phi)}d_{K
m}^{J}(\Theta)\exp{(-im\varphi)},\label{WF}\end{equation}
\begin{equation}
\int\limits_0^{2\pi}\rmd
\Phi\int\limits_0^{\pi}\sin\Theta\rmd\Theta
\int\limits_0^{2\pi}\rmd \varphi D_{K m}^{J\lambda *}D_{K'
m'}^{J'\lambda'}=\delta_{JJ'}\delta_{KK'}\delta_{mm'}
\delta_{\lambda\lambda'}\label{DD}.\end{equation}

We define the hyperspherical Coulomb spheroidal (HSCS) basis on
the 5-dimensional hypersphere
\begin{equation}
\Omega=
\{t,\vartheta,\Phi,\Theta,\varphi\}=\{\xi,\eta,\Phi,\Theta,\varphi\},
\label{Omeg5}
\end{equation}
\begin{equation}\fl
\rmd\Omega=t^2 \sin\vartheta \rmd
t\rmd\vartheta\sin\Theta\rmd\Phi\rmd\Theta\rmd\varphi
=\frac{(t_1+t_2)^3}{8}(\xi^2-\eta^2)\rmd \xi\rmd\eta \sin\Theta
\rmd\Phi\rmd\Theta\rmd\varphi \label{dOmeg5}\end{equation} as the
system of the common eigenfunctions of the operators $\bi{J}^2,
J_3, {J'}_3^2$, $P$ and the Coulomb-two-centre Hamiltonian $h$
\eref{h} depending on the parameter $\rho$ :
\begin{equation}
\Phi^{JK\lambda}_{jm}(\rho|\Omega)
= D_{K m}^{J\lambda}(\Phi,\Theta,\varphi)\phi_{jm}(\rho|\xi,\eta).
\label{Phi}
\end{equation}

Due to the completeness and the orthonormality  of the system
$\{\Phi^{JK\lambda}_{jm}\}$ \eref{Phi} the three-body wave
function $\tilde\Psi$ \eref{tPsi} with fixed $J,K,\lambda$ is
presented in the form of the HSCS-decomposition
\begin{equation}
\tilde\Psi^{JK\lambda}(\rho,\Omega)=\rho^{5/2}(1+t^2)^{-3/2}
\Psi^{JK\lambda}(\rho,\Omega) =\sum_{jm} f^{JK\lambda jm }(\rho)
\Phi^{JK\lambda}_{jm}(\rho|\Omega), \label{tPsi2}
\end{equation}
where the sum by $j$ includes both the discrete spectrum and the
continuum (see \eref{compl}). The coefficients of the
decomposition \eref{tPsi2}, the radial functions
$f^{JK\lambda}_{jm}(\rho)$, are given by the integral
\begin{equation}
f^{JK\lambda jm}(\rho)= \int \Phi^{JK\lambda *}_{jm}(\rho|\Omega)
\tilde\Psi^{JK\lambda}(\rho|\Omega)\rmd\Omega .
\label{f}\end{equation}

\section{$S$-matrix and radial functions}

\subsection{Basis functions of the discrete spectrum at large $\rho$}

Every eigenfunction of the discrete spectrum of $h$ \eref{h} at
large $\rho$ tends to the wave function of the bound state of the
atom (1,3) ($\alpha=1$) or the atom (2,3) ($\alpha=2$). The
detailed analysis \cite{KPS} shows that the limiting atomic states
are the states with the definite parabolic quantum numbers in the
rotating frame. Thus, the basis functions \eref{phi} can be
enumerated by the indices $\alpha$ (the number of the nucleus
forming the atom), $n$ (the principal quantum number), $s$ (the
first parabolic quantum number) and $m$ (the absolute value of the
projection of the atomic orbital momentum on the axis $\bi{R}$):
\begin{equation}
\phi_{n_{\xi}n_{\eta}m}(\rho|\xi,\eta)= \phi_{\alpha n s
m}(\rho|\xi,\eta).\label{ansm}
\end{equation}
The formulae determining the one-to-one correspondence
\begin{equation} \{n_{\xi},n_{\eta},m\} \leftrightarrow
\{\alpha,n,s,m\} \label{1to1}
\end{equation} for different cases are presented in the monograph
\cite{KPS}.

To obtain the factor connecting the asymptotics of $\phi_{\alpha n
s m}(\rho|\xi,\eta)$ and the normalized atomic function in the
region \begin{equation}\rho\to\infty,\qquad r_{\alpha}=O(1)
\label{reg}\end{equation} one have to use the asymptotical formula
\begin{equation}
|{\bi{t}}-{\bi{t}}_1|=
r_{\alpha}(2/\mu_{\alpha})^{1/2}\mu/\rho+O(\rho^{-2})
\end{equation} and relations \eref{norma}, \eref{38}
defining the normalization of basis functions. The result is
\begin{equation} \phi_{\alpha n s m}(\rho|\xi,\eta)
\Bigr|_{\rho\to\infty}=
(\mu_{\alpha}\rho^2/2\mu^2)^{3/4}\phi_{\alpha n s
m}^{(at)}({r}_{\alpha},\vartheta_{\alpha}).\label{as-phi}
\end{equation}
Here $\phi_{\alpha n s m}^{(at)} ({r}_{\alpha},
\vartheta_{\alpha})$ is the normalized atomic function of the atom
$\alpha$ without the factor $\exp{(\pm im\varphi)}/\sqrt{2\pi}$
\begin{equation}
\int\limits_0^{\infty} r_{\alpha}^2\rmd
r_{\alpha}\int\limits_o^\pi \sin\vartheta_{\alpha}
\rmd\vartheta_{\alpha} |\phi_{\alpha n s m}^{(at)}({r}_{\alpha},
\vartheta_{\alpha})|^2 =1,\label{norm}
\end{equation}
$\vartheta_{\alpha}$ is the angle between $\bi{r}_{\alpha}$ and
$\bi{R}$ . The atomic function is factorized in the parabolic
coordinates, but we conserve the notation $\phi_{\alpha n s
m}^{(at)}({r}_{\alpha},\vartheta_{\alpha})$ for convenience.

 Using equation \eref{as-phi} we
obtain the asymptotics of the HSCS basis function
$\Phi^{JK\lambda}_{\alpha n s m}$ \eref{Phi} in the form
\begin{equation}
\Phi^{JK\lambda}_{\alpha n s m}(\rho|\xi,\eta,\Phi,\Theta,\varphi)
\Bigr|_{\rho\to\infty}=(\mu_{\alpha}\rho^2/2\mu^2)^{3/4}
F^{JK\lambda}_{\alpha n s m}(\Phi,\Theta,\varphi,{r}_{\alpha},
\vartheta_{\alpha}),\label{AsPhi}
\end{equation} where
\begin{eqnarray}
\fl F^{JK\lambda}_{\alpha n s m}
(\Phi,\Theta,\varphi,{r}_{\alpha},\vartheta_{\alpha}) = D_{K
m}^{J\lambda}(\Phi,\Theta,\varphi) \phi_{\alpha n s
m}^{(at)}({r}_{\alpha},\vartheta_{\alpha})\nonumber\\
\lo= A^J_{Km} \left[ d_{-K m}^{J}(\Theta) \exp(-im\varphi)+\lambda
(-1)^{J+m}
d_{-K -m}^{J}(\Theta) \exp(im\varphi) \right]\nonumber\\
\times\exp\left(iK\Phi\right)\phi_{\alpha n s
m}^{(at)}({r}_{\alpha},\vartheta_ {\alpha}) . \label{F}
\end{eqnarray}

It is seen that $F^{JK\lambda}_{\alpha n s m}$ is the combination
of two atomic functions with opposite projections of the orbital
momentum on the rotating axis ${\bi{R}}$ with coefficients
depending on $\Phi$ and $\Theta$. Functions $F^{JK\lambda}_{\alpha
n s m}$ corresponding to the same atom ($\alpha=\alpha'$) satisfy
the orthogonality condition
\begin{eqnarray}\fl
\int\limits_0^{2\pi} \rmd\Phi
\int\limits_0^{\pi}\sin{\Theta}\rmd\Theta
\int\limits_0^{2\pi}\rmd\varphi \int\limits_0^{\infty}{r}^2
_{\alpha}\rmd{r}_{\alpha}
\int\limits_0^{2\pi}\sin{\vartheta}\rmd\vartheta_{\alpha}
F^{JK\lambda}_{\alpha n s m}F^{J'K'\lambda'}_{\alpha n' s' m'}
\nonumber\\=
\delta_{JJ'}\delta_{KK'}\delta_{\lambda\lambda'}\delta_{mm'}
\delta_{nn'}\delta_{ss'}\label{FF}.
\end{eqnarray}

\subsection{S-matrix in the ($\alpha n s m$)-representation}

The product of $F^{JK\lambda}_{\alpha n s m}$ \eref{F} and the
ingoing (outgoing) wave
\begin{equation}
F^{JK\lambda}_{\alpha n s m}
(\Phi,\Theta,\varphi,{r}_{\alpha},\vartheta_{\alpha})\cdot
R_{\alpha}^{-1}{\exp{[\pm i(k_{\alpha n} R_{\alpha}-\gamma_{\alpha
n} )]}} \label{wave}\end{equation} presents the state of the
colliding atom and the remaining particle ((1,3)+2 for $\alpha=1$
or (2,3)+1 for $\alpha=2$) at large distance $R_{\alpha}$ between
them. Here $k_{\alpha n}$ is the relative momentum
\begin{equation}k_{\alpha n}=\sqrt{2M_{\alpha}(E-E_{\alpha n})}=
M_{\alpha}v_{\alpha n}, \label{k}\end{equation} $E_{\alpha n}$ is
the energy of the isolated atom $\alpha$, the logarithmic phase
$\gamma_{\alpha n}$ is given by
\begin{equation}\gamma_{\alpha n}=
\gamma_{\alpha n}(R_{\alpha})=(Z_{\alpha}-1)
Z_{3-\alpha}M_{\alpha}k^{-1}_{\alpha n} \log{2k_{\alpha
n}R_{\alpha}} .\label{gamma}\end{equation} The asymptotics of any
three-body wave function at energy below the threshold of the
three-particle breakup is presented as the combination of the
products \eref{wave}.

Consider the three-body system with fixed $J,K,\lambda$. These
indices are omitted hereinafter. According to the general theory
\cite{Landau} for the definition of $S$-matrix we have to consider
the three-body wave function which contains the single ingoing
wave in the input channel $\{\alpha,n,s,m\}$ and the outgoing
waves in all open channels $\{\alpha',n',s',m'\}$
\begin{eqnarray} \fl\Psi_{\alpha n s
m}\Bigr|_{R\to\infty}=F_{\alpha n s m }\frac{\exp{[-i(k_{\alpha
n}R_{\alpha}-\gamma_{\alpha n})]}}
{R_{\alpha}\sqrt{v_{\alpha n}}}\nonumber\\
 -\sum_{\alpha'n's'm'} \tilde S^{\alpha'n's'm'}_{\alpha n
s m}F_{\alpha'n's'm'}\frac{\exp{[i(k_{\alpha'
n'}R_{\alpha'}-\gamma_{\alpha' n'})]}}
{R_{\alpha'}\sqrt{v_{\alpha' n'}}}.\label{Asss}
\end{eqnarray}
Here the coefficients $\tilde S^{\alpha' n' s'm'}_{\alpha n sm}$
are the matrix elements of the operator
\begin{equation}\hat{\tilde
S}$=$\hat{S}\hat{I},\label{tilds}\end{equation} where $\hat{S}$ is
the $S$-matrix, and $\hat{I}$ is the inversion operator which
changes the sign of the relative coordinate ${\bi{R}}_{\alpha}$ at
fixed $\bi{r}_{\alpha}$. It acts on any function of  $\Phi$,
$\Theta$, $\varphi$, $\vartheta_{\alpha}$ as follows
\begin{equation}
\hat{I}Y\left(\Phi,\Theta,\varphi,\vartheta_{\alpha}\right)=
Y\left(\Phi+\pi,\pi-\Theta,-\varphi,\pi-\vartheta_{\alpha}\right).
\label{I}\end{equation}

The matrix elements of $S$-matrix in any representation can be
obtained from $\tilde S^{\alpha' n' s'm'}_{\alpha n sm}$ in the
regular way with the use of relations \eref{tilds}, \eref{I} and
the transformation matrix connecting the set
$\{F^{JK\lambda}_{\alpha n s m}\}$ \eref{F} with the corresponding
new set.

\subsection{The asymptotics of radial functions in terms of $\tilde S^{\alpha'n'
s'm'}_{\alpha n sm}$.}

According to \eref{tPsi2}, \eref{f} the wave function
$\Psi_{\alpha n sm}$ defined by the asymptotical condition
\eref{Asss} can be presented in the form of the HSCS-decomposition
\begin{equation}\Psi_{\alpha n sm} =\rho^{-5/2}(1+t^2)^{3/2}\sum_{jm'}
f^{jm'}_{\alpha n sm}(\rho)\Phi_{jm'}(\rho|\Omega), \label{Psi-a}
\end{equation}
where the radial functions are are given by
\begin{equation}
f^{jm'}_{\alpha n sm}(\rho)= \int
\Phi_{jm'}^{*}(\rho|\Omega)\rho^{5/2}(1+t^2)^{-3/2} \Psi_{\alpha
nsm}(\rho|\Omega)\rmd\Omega. \label{ff}\end{equation} Using this
relation and the asymptotical formula \eref{Asss} for
$\Psi_{\alpha n sm}$ we can express the asymptotics of
$f^{jm'}_{\alpha nsm}(\rho)$ in terms of the matrix elements
$\tilde S^{\alpha' n' s'm'}_{\alpha n sm}$. To this end we have to
rewrite the expression \eref{Asss} in the hyperspherical
coordinates using the relation
\begin{equation}R_{\alpha}=(2M_{\alpha})^{-1/2}\rho+O(\rho^{-1})
\label{rho} \end{equation} which is valid in the asymptotical
region \eref{reg}. As a result we have
\begin{eqnarray}\fl
\Psi_{\alpha n sm}(\rho|\Omega)\Bigr|_{\rho\to\infty}=F_{\alpha n
sm} 2^{1/4}M_{\alpha}^{3/4} \frac{\exp{[-i(q_{\alpha
n}\rho-\bar\gamma_{\alpha n})]}} {q_{\alpha
n}^{1/2}\rho}\nonumber\\ - \sum_{\alpha'n's'm'} \tilde S^{\alpha'
n' s'm'}_{\alpha n sm}F_{\alpha' n' s'm'}2^{1/4}M_{\alpha'}^{3/4}
\frac{\exp{[i(q_{\alpha' n'}\rho-\bar\gamma_{\alpha' n'}
)]}}{q_{\alpha' n'}^{1/2}\rho },\label{Ass}
\end{eqnarray} where
\begin{equation}
q_{\alpha n}=\sqrt{E-E_{\alpha n}}, \label{q}\end{equation}

\begin{equation}\bar\gamma_{\alpha n}=
\bar\gamma_{\alpha n}(\rho)=(Z_{\alpha}-1)
Z_{3-\alpha}M_{\alpha}k^{-1}_{\alpha n} \log{2q_{\alpha n}\rho}
.\label{gama}\end{equation}

The substitution of $\Psi_{\alpha n sm}(\rho|\Omega)$ by the
asymptotic expression \eref{Ass} in equation \eref{ff} and the
integration with taking into consideration \eref{AsPhi} and
\eref{FF} lead to the asymptotical formulae for radial functions
$f^{\alpha'n's'm'}_{\alpha n sm}(\rho)$ where $\{\alpha n sm\}$ is
the index of the input channel. For indices
$\{jm'\}=\{\alpha'n's'm'\}$ corresponding to open channels we
obtain
\begin{eqnarray}\fl
f^{\alpha'n's'm'}_{\alpha n sm}(\rho)
\Bigr|_{\rho\to\infty}=2(M\mu)^{3/4}q_{\alpha n}^{-1/2}\left\{
\delta_{\alpha\alpha'}\delta_{nn'}\delta_{ss'} \delta_{mm'}
\exp[-i(q_{\alpha n}\rho-\bar\gamma_{\alpha n})]\right.\nonumber\\
-\left. \left(q_{\alpha n}/q_{\alpha' n'}\right)^{1/2}\tilde
S^{\alpha' n' s'm'}_{\alpha n sm} \exp[i(q_{\alpha'
n'}\rho-\bar\gamma_{\alpha n})]\right\}. \label{f-as}
\end{eqnarray}

For $\{j,m'\}$ corresponding to closed channels and the continuous
spectrum of the Hamiltonian $h$ \eref{h} the result is
\begin{equation}f^{jm'}_{\alpha n sm}(\rho)\Bigr|_{\rho\to\infty}=0.
\label{f-as2}
\end{equation}

\subsection{The coordination with the boundary
conditions of the scattering problem}

The relations \eref{f-as} and \eref{f-as2} show that the radial
function $f^{\alpha'n's'm'}_{\alpha n sm}(\rho)$ does not vanish
at $\rho\to\infty $ only if its indices ${\alpha'n's'm'}$
correspond to one of open channels. In that case its asymptotics
contains the single matrix element $\tilde S^{\alpha' n'
s'm'}_{\alpha n sm}$. In other words, the function describing the
motion of the non-interacting fragments in the definite open
channel (the definite term of the expression \eref{Asss}) is the
asymptotics of the single term of the decomposition \eref{Psi-a}.
Other terms of \eref{Psi-a} do not make contribution to that
function. Thus, the HSCS expansion \eref{Psi-a} is coordinated
with the boundary conditions of the scattering problem.

That is not right for the traditional BO approach where $R$ is an
adiabatic parameter. The reason is that the BO basis functions
form an orthogonal system on the surface $R=const$, so that
surface appears in the integral for the radial function which
replaces \eref{ff} in the BO case. For the calculation of that
integral one has to to use for $R_{\alpha}$ the expansion in
powers of $R^{-1}$. But that expansion, in contrast to the
expansion in powers of $\rho^{-1}$ \eref{rho}, contains the
non-vanishing term depending on the atomic coordinate
$\bi{r}_{\alpha}$:
\begin{equation} R_{\alpha}\Bigr|_{R\to\infty}=R+(-1)^{\alpha}m_3
(m_{\alpha}+m_3)^{-1}R^{-1} \bi{r}_{\alpha}\cdot\bi{R}+O\left(
R^{-1}\right) . \label{Ral}
\end{equation}
Consequently, the asymptotic formula for the ingoing (outgoing)
wave $\exp(\pm ik_{\alpha n} R_{\alpha})$ contains the factor
\begin{equation}\exp\left[\pm ik_{\alpha n} {m_3}
{(m_{\alpha}+m_3)^{-1}} {R^{-1}} \bi{r}_{\alpha}\cdot\bi{R}
\right]. \label{expr}
\end{equation}
The matrix elements of that factor  differs from zero for all
states of the atom $\alpha$ including the excited states
corresponding to closed channels. So the radial function does not
vanish at $R\to\infty$ for all states with $\alpha'=\alpha$. It
means that all terms of the traditional BO decomposition with
given $\alpha$ make contribution to the wave functions of all open
channels with that $\alpha$.

\subsection{The real radial functions}

In the numerical calculations it is suitable to use instead of the
complex functions $\Psi_{\alpha n s m}$ \eref{Psi-a}, \eref{Asss}
their real linear combinations $G_{\alpha n sm}$ which contain
$\sin(k_{\alpha n}R_{\alpha}-\gamma_{\alpha n}-\pi J/2)$ in the
input channel and $\cos(k_{\alpha' n'}R_{\alpha'}- \gamma_{\alpha'
n'}-\pi J/2)$ in all open channels. The HSCS representation for
$G_{\alpha n sm}$ has the form
\begin{equation}G_{\alpha n sm} =\rho^{-5/2}(1+t^2)^{3/2}\sum_{jm'}
g^{jm'}_{\alpha n sm}(\rho)\Phi_{jm'}(\rho|\Omega), \label{G}
\end{equation} where the asymptotics of the radial function
$g^{jm'}_{\alpha n sm}(\rho)$ with indices
$\{jm'\}=\{\alpha'n's'm'\}$ corresponding to open channel is given
by
\begin{eqnarray}\fl
g^{\alpha'n's'm'}_{\alpha n sm}(\rho) \Bigr|_{\rho\to\infty}=
\delta_{\alpha\alpha'}\delta_{nn'}\delta_{ss'} \delta_{mm'}
\sin(q_{\alpha n}\rho-\bar\gamma_{\alpha n}-\pi J/2)\nonumber\\
- \left(q_{\alpha n}/q_{\alpha' n'}\right)^{1/2}K^{\alpha' n'
s'm'}_{\alpha n sm} \cos(q_{\alpha' n'}\rho-\bar\gamma_{\alpha
n}-\pi J/2). \label{g-as}
\end{eqnarray}
For $\{j,m'\}$ corresponding to closed channels and the continuous
spectrum we have
\begin{equation}g^{jm'}_{\alpha n sm}(\rho)\Bigr|_{\rho\to\infty}=0.
\label{g-as2}
\end{equation}
The relation between $K^{\alpha' n' s'm'}_{\alpha n sm}$
\eref{g-as} and $\tilde S^{\alpha' n' s'm'}_{\alpha n sm} $
\eref{f-as}  can be written 
as follows
\begin{equation}
\textsf{K}=i\left[\textsf{1}-(-1)^J\tilde \textsf{S}\right]
\left[\textsf{1}+(-1)^J\tilde \textsf{S}\right]^{-1}, \label{K}
\end{equation}

\begin{equation}
\tilde\textsf{S}=(-1)^J\left[\textsf{1}-i\textsf{K}\right]^{-1}
\left[\textsf{1}+i\textsf{K}\right], \label{S}\end{equation} where
$\tilde\textsf{S}$ and $\textsf{K} $ are the matrices with matrix
elements
\begin{equation}
\tilde S_{fi}= \tilde S^f_i,\quad K_{fi}= K^f_i,\quad f=\{\alpha'
n' s' m'\},\quad    i=\{\alpha n s m\}. \label{SK}\end{equation}
The relation between functions $\Psi_{\alpha nsm}$ and $G_{\alpha
nsm}$ is given by
\begin{equation}
G_i=2^{-1}(M\mu)^{-3/4}q_i^{1/2}\exp[i\pi(J+1)/2]\sum_f \Psi_f
\left[\textsf{1}+(-1)^J\tilde
\textsf{S}\right]^{-1}_{fi}.\label{GPsi}\end{equation}

\section{The scattering problem for the radial system}

For the calculation of $K$-matrix it necessary to solve the system
of equations for the radial functions which follows from the
Schr\"{o}dinger equation. To obtain that system in the simplest
form it is suitable to present the AHS Hamiltonian $\tilde h^{\rm
a\rmd}$  \eref{had} as the sum
\begin{equation}
\tilde h^{\rm a\rmd}=\rho^{-2}(1+t^2)h(1+t^2)+W,\end{equation}
where the additional potential $W(\rho,\bi{t})$ has no
singularities in the coalescence points. It is given by the
formulae
\begin{equation}\fl W=V+\frac{(1+t^2)^2}{\rho^2}\left[\frac{\tilde
Z_1} {|\bi{t}-\bi{t}_1|}+\frac{\tilde Z_2}{|\bi{t}-\bi{t}_2|}
\right]=\frac{1}{\rho}\left[w^{(12)}+w^{(13)}+w^{(23)}\right],
\label{trV} \end{equation}
\begin{equation}\eqalign{
\fl w^{12}=Z_1Z_2(2M)^{1/2}(1+t^2)^{1/2},\\
\fl w^{(13)}=\frac{Z_1(2\mu)^{1/2}(1+t^2)^{1/2}}{|\bi{t}
-\bi{t}_1|} \left[
\left(\frac{1+t^2}{1+t_1^2}\right)^{3/2}-1\right]
\stackrel{\bi{t}\to\bi{t_1}}
{\rightarrow}3Z_1\sqrt{2(\mu-M_{13})},\\
\fl
w^{(23)}=\frac{Z_2(2\mu)^{1/2}(1+t^2)^{1/2}}{|\bi{t}-\bi{t}_2|}
\left[ \left(\frac{1+t^2}{1+t_2^2}\right)^{3/2}-1\right]
\stackrel{\bi{t}\to\bi{t_2}}
{\rightarrow}3Z_2\sqrt{2(\mu-M_{23})}.}\label{ww}\end{equation}

The three-body Hamiltonian $\tilde H$ \eref{tH} takes the form
\begin{equation}\fl
\tilde H=-\frac{\partial^2}{\partial\rho^2}+\frac{1}{\rho^2}\left[
(1+t^2)h(1+t^2) +(1+t^2)(\bi{J}^2-2\bi{J}\cdot\bi{l})
+\frac{3}{4}\right]+ W.\label{e25}\end{equation} The substitution
of $\tilde \Psi^{JK\lambda}$ by the decomposition \eref{tPsi2} in
the Schr\"{o}dinger equation \eref{tSchr}, \eref{e25} and the
projecting  onto the state $\Phi_{im}^{JK\lambda}(\rho| \Omega)$
lead to the system of the differential equations for radial
functions
\begin{eqnarray}\fl
\left(-\frac{d^2}{d\rho^2}+\frac{3}{4\rho^2}-E\right)f^{im}(\rho)
+\sum_j\left[\left(
2Q^m_{ij}\frac{d}{d\rho}+\frac{dQ^m_{ij}}{d\rho}+P^m_{ij}+
U^m_{ij}+W^m_{ij}+\frac{R^m_{ij}}{\rho^2}\right) f^{jm}(\rho)
\right.
\nonumber\\
\left.+\frac{1}{\rho^2}\left(T_{im,jm-1}f^{jm-1}(\rho)+
T_{jm+1,im}f^{jm+1}(\rho)\right)\right]
=0,\label{sys}\end{eqnarray} where the sum by $j$ includes both
the discrete spectrum and the continuum (see \eref{compl}). The
indices $J,K,\lambda $ corresponding to the integrals of motion
are omitted.

The matrix elements $P^m_{ij}$, $Q^m_{ij}$, $R^m_{ij}$,
$T_{im,jm'}$, $U^m_{ij}$, $W^m_{ij}$ in \eref{sys} are the
two-dimensional integrals containing the factorized solutions of
the CTC problem $\phi_{jm}(\rho|\xi,\eta)$ \eref{phi}:
\begin{equation}\eqalign{
P^m_{ij}(\rho) = P^m_{ji}(\rho) =  \biggl< {{\partial
}\over{\partial \rho}} \phi_{im} | {{\partial }\over{\partial
\rho}} \phi_{jm} \biggr>, \\
Q^m_{ij}(\rho)  = -Q^m_{ji}(\rho) = \biggl< {{\partial
}\over{\partial \rho}} \phi_{im} | \phi_{jm}
\biggr>, \\
R^m_{ij}(\rho)=R^m_{ji}(\rho)=[J(J+1)-2m^2]\langle
\phi_{im} |(1+t^2)| \phi_{jm}\rangle, \\
T_{im,jm'}(\rho)= [J(J+1)-mm']^{1/2}(1+\delta_{m1})^{1/2}\times\\
\times \langle\phi_{im}|(1+t^2)( -{\partial }/{\partial \vartheta}
+ m'\cot\vartheta)|
\phi_{jm'}\rangle, \\
U^m_{ij}(\rho)= \frac{1}{\rho^2}
\langle\phi_{im}|(1+t^2)h_m(1+t^2)|\phi_{jm}\rangle,\\
W^m_{ij}(\rho)=\langle\phi_{im}|W|\phi_{jm}\rangle=\frac{1}{\rho}
\langle\phi_{im}|w^{(12)}+w^{(13)}+w^{(23)}|\phi_{jm}\rangle,
}\label{MTR}\end{equation} the operator $h_m$ is given by the
expression
\begin{eqnarray}\fl
h_m=\frac{1}{{2\pi}}\int\limits^{2\pi}_{0}e^{-im\varphi}he^{im\varphi}
\rmd\varphi=\nonumber\\=-\frac{1}{t^2}\frac{\partial}{\partial
t}t^2\frac{\partial}{\partial t}- {{1}\over{t^2\sin \vartheta}}
\left[ {{\partial }\over{\partial \vartheta}} \sin \vartheta
{{\partial }\over{\partial \vartheta}} +{{m^2}\over{\sin
\vartheta}}\right] -\frac{\tilde Z_1}
 {|\bi{t}-\bi{t}_1|}-\frac{\tilde Z_2}{|\bi{t}-\bi{t}_2|}.
\end{eqnarray}

It is evident that the real radial functions $g^{jm'}(\rho)$
\eref{G} also satisfy the system \eref{sys} as well as the
coefficients of the HSCS decomposition of any solution of the
Schr\"{o}dinger equation \eref{tSchr}.

We did not study the asymptotics of the matrix elements at large
$\rho$ in details as it is not necessary for the statement of the
scattering problem. However, the principal term of the asymptotics
of the diagonal matrix element $U^m_{ii}$ is obtained without any
complicated calculations:
\begin{equation}
U^m_{ii}(\rho)\Bigr|_{\rho\to\infty}=E_{\alpha n},\qquad
i=\{\alpha ns\}.\label{U} \end{equation} Other matrix elements in
\eref{sys} vanish at $\rho\to\infty$.

To find the matrix $K^{\alpha' n' s'm'}_{\alpha n sm}$ \eref{g-as}
one has to solve the system \eref{sys} for the real radial
functions $g_{\alpha nsm }^{jm'}(\rho)$ for all sets $\{\alpha
nsm\}$ (the input channel) corresponding to open channels. The
solutions have to satisfy the boundary conditions at $\rho=0$
\begin{equation}g_{\alpha nsm}^{jm'}(0)=0,\end{equation}
which follow from the finiteness of the CTB wave function $G$
\eref{G}, and the asymptotical conditions \eref{g-as},
\eref{g-as2} at $\rho\to\infty$. The matrix elements $\tilde
S^{\alpha' n' s'm'}_{\alpha n sm}$ are expressed in terms of
$K^{\alpha' n' s'm'}_{\alpha n sm}$ in accordance with \eref{S}.
The matrix elements of $S$-matrix can be expressed in terms of
$\tilde S^{\alpha' n' s'm'}_{\alpha n sm}$ with the help of
relation \eref{tilds}.

\section{Concluding remarks}

The  suggested  HSCS representation is coordinated with the
boundary conditions of the scattering problem like the AHS one but
the HSCS basis functions are essentially simpler. The combination
of the simplicity and the coordination mentioned is achieved at
the price of the deviation from the adiabatic idea and the
corresponding complication of the radial equations. Indeed, the
CTC Hamiltonian $h$ \eref{h} generating our basis functions
differs from the AHS Hamiltonian $\tilde h^{\rm a\rmd}$
\eref{had1} by both the potential energy and the kinetic one.
Therefore the radial system \eref{sys} contains the non-diagonal
matrices $U^m_{ij}(\rho)$ and $W^m_{ij}(\rho)$ \eref{MTR} instead
of the diagonal matrix of $\tilde h^{\rm a\rmd}$ in the AHS
approach. That complication, however, looks insignificant as
compared with the advantages resulting from the properties of the
HSCS expansion. First, all calculations are simplified. Second,
the avoided crossings of the AHS terms are changed by the exact
crossings of the CTC terms. Third, the use of the well-known and
relatively simple CSF allows to perform the precise calculations
in a wide range of $\rho$ including extremely large values which
are essential for the calculation of $S$-matrix. These theoretical
reasons need, of course, the practical verification.

The essential difference between  HSCS and AHS expansions is that
the first one includes both discrete and continuous spectra while
the second one has a pure discrete spectrum. In the AHS approach
(as well as in the approach of papers \cite{T1} - \cite{T5}) every
basis function at large $\rho$ is localized in the vicinity of the
coalescent point where it coincides with the wave function of the
bound state of the corresponding atom. Thus, that basis is not
suitable for the representation of the CTB wave function at
energies above the threshold of the three-particle breakup. In
contrast to the AHS basis, the HSCS one includes the functions
which are not localized in the vicinities of the coalescent points
at any $\rho$. That allows to hope that the HSCS basis can be used
for the adequate representation of the CTB wave function at
energies above the threshold mentioned.

The method suggested, by analogy with the adiabatic hyperspherical
approach, can be named "the diabatic hyperspherical approach" as
the avoided crossings of the AHS terms correspond to the exact
crossings of the CTC terms. In this connection it is interesting
to develop the semiclassical version of the HSCS approach.

It is interesting also to apply the HSCS expansion to the
calculation of the bound states of the CTB systems. That problem
is simpler than the scattering one as all radial functions vanish
at $\rho\to\infty$.

The applied three-pole coordinate system on the sphere
$\rho=const$ which is connected with the elliptic coordinate
system on the plane by means of the stereographic projection may
be interesting for other three-body problems in the hyperspherical
approach as it presents the natural way to take into consideration
three coalescent points of the three-body system.

\ack The author is grateful to V.V.Gusev for helpful discussions.

\end{document}